\documentclass[aps,reprint,superscriptaddress,longbibliography,prb]{revtex4-2}
\usepackage{graphicx}
\usepackage{amsmath}
\DeclareMathOperator\arctanh{arctanh}

\usepackage{amsfonts}
\usepackage{comment}
\usepackage{bbm}
\usepackage{bm}
\usepackage{comment}
\usepackage{esint}
\usepackage{cancel}
\usepackage[dvipsnames]{xcolor}
\usepackage[unicode=true, colorlinks=true, citecolor={blue!80!black}, urlcolor={blue!50!black}, linkcolor = {blue!80!black}]{hyperref}
\usepackage{microtype}
\usepackage[normalem]{ulem}
\usepackage{mathrsfs}
\usepackage{multirow}
\usepackage{array}
\usepackage{lipsum}
\newcolumntype{C}[1]{>{\centering\arraybackslash}p{#1}}

\newcommand{\abs}[1]{\left\vert#1\right\vert}
\newcommand{\ket}[1]{\left\vert#1\right\rangle}
\newcommand{\bra}[1]{\left\langle#1\right\vert}

\newcommand{\up}{\uparrow}
\newcommand{\down}{\downarrow}

\graphicspath{{../figures/}}

\begin{document}
\title{
Tunneling of fluxons via a Josephson resonant level
}
\author{T. Vakhtel}
\affiliation{Instituut-Lorentz, Universiteit Leiden, P.O. Box 9506, 2300 RA Leiden, The Netherlands}
\author{P. D. Kurilovich}
\affiliation{Departments of Applied Physics and Physics, Yale University, New Haven, CT 06520, USA}
\author{M. Pita-Vidal}
\affiliation{QuTech, Delft University of Technology, Delft 2628 CJ, The Netherlands}
\affiliation{Kavli Institute for Nanoscience, Delft University of Technology, Delft 2628 CJ, The Netherlands}
\author{A. Bargerbos}
\affiliation{QuTech, Delft University of Technology, Delft 2628 CJ, The Netherlands}
\affiliation{Kavli Institute for Nanoscience, Delft University of Technology, Delft 2628 CJ, The Netherlands}
\author{V. Fatemi}
\affiliation{School of Applied and Engineering Physics, Cornell University, Ithaca, NY 14853, USA}
\author{B. van Heck}
\affiliation{Dipartimento di Fisica, Sapienza Università di Roma, Piazzale Aldo Moro 2, 00185 Rome, Italy}

\date{\today}
\begin{abstract}
Fluxons in a superconducting loop can be coherently coupled by quantum phase slips occurring at a weak link such as a Josephson junction.
If Cooper pair tunneling at the junction occurs through a resonant level, $2\pi$ quantum phase slips are suppressed, and fluxons are predominantly coupled by $4\pi$ quantum phase slips.
We analyze this scenario by computing the coupling between fluxons as the level is brought into resonance with the superconducting condensate.
The results indicate that the $4\pi$-dominated regime can be observed directly in the transition spectrum for circuit parameters typical of a fluxonium qubit.
We also show that, if the inductive energy of the loop is much smaller than the plasma frequency of the junction, the low-energy Hamiltonian of the circuit is dual to that of a topological superconducting island.
These findings can inform experiments on bifluxon qubits as well as the design of novel types of protected qubits.
\end{abstract}

\maketitle

\section{Introduction}

The inductively shunted Josephson junction plays an important role in the field of superconducting quantum devices~\cite{clarke2008superconducting,blais2021circuit}.
The inductive link changes the topology of the circuit from that of an island to that of a loop, removing the $2e$ charge quantization associated with a superconducting island.
The charge sensitivity of the device is exchanged for its flux sensitivity~\cite{koch2009charging}, which is exploited in the design and operation of the fluxonium qubit~\cite{manucharyan2009fluxonium, nguyen2019high,nguyen2022blueprint,bao2022fluxonium, somoroff2023millisecond}.
Furthermore, a large shunting inductance suppresses the sensitivity to flux noise, as recently demonstrated in the blochnium qubit~\cite{pechenezhskiy2020superconducting}.
For this reason, the inductive shunt is a common feature of noise-protected qubit designs~\cite{gyenis2021moving}.

\begin{figure}[t!]
    \centering
    \includegraphics[width=\columnwidth]{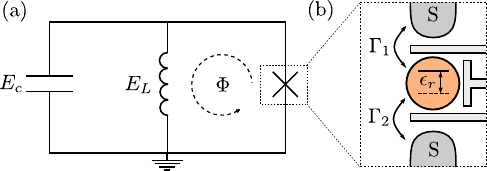}
    \caption{(a) Circuit of the inductively shunted Josephson junction. (b) A junction realized by a resonant level with a tunable energy $\epsilon_r$ and Cooper pair tunneling rates $\Gamma_1$ and $\Gamma_2$.
    }
    \label{fig:layout}
\end{figure}

The minimal circuit that models this class of superconducting devices is simple: it consists of an inductor, a capacitor and a Josephson element connected in parallel [Fig.~\ref{fig:layout}(a)].
The inductor and the Josephson junction form a loop through which an applied magnetic flux $\Phi$ is threaded.
The circuit supports persistent current states, also known as \emph{fluxons}, in which the superconducting phase winds by an integer multiple $m$ of $2\pi$ when circling the loop~\cite{matveev2002}.
Fluxons are coupled by quantum phase slips occurring at the Josephson junction~\footnote{In principle, phase slips may also occur at other points in the loop, through the inductor. We neglect this possibility, which is analyzed in Ref.~\cite{manucharyan2012}}, which change $m$ by an integer $\Delta m$~(see Fig.~\ref{fig:landscape}).

In a typical Josephson element, e.g. in a tunnel junction, the amplitude of $2\pi$ quantum phase slips ($\Delta m=1$) is much larger than that of $4\pi$ quantum phase slips ($\Delta m=2$).
However, if Cooper pair tunneling across the Josephson element is resonant -- a type of weak link we call the Josephson resonant level -- $2\pi$ quantum phase slips are suppressed~\cite{averin1999coulomb,ivanov1998coulomb,averin1999quantum,vakhtel2023quantum} and $4\pi$ quantum phase slips become the dominant coupling between fluxons.
The bifluxon qubit proposal~\cite{kalashnikov2020bifluxon} achieves resonant tunneling using as a Josephson element a series of two (almost) identical tunnel junctions separated by a small superconducting island tuned (close) to a charge degeneracy point.
Alternatively, resonant tunneling can also occur in a semiconductor junction, via an isolated energy level forming in a quantum dot~\cite{beenakker1992resonant,beenakker1992three,devyatov1997resonant}, as represented in Fig.~\ref{fig:layout}(b).
In the latter system, experiments have demonstrated the drastic suppression of $2\pi$ quantum phase slips close to resonance~\cite{bargerbos2020observation,kringhoj2020suppressed}, but not yet the occurrence of the regime dominated by $4\pi$ quantum phase slips~\cite{vakhtel2023quantum}.

In this paper, motivated by these experimental developments, we study in detail the energy spectrum of an inductively shunted junction with a Josephson coupling mediated by a single energy level [Fig.~\ref{fig:layout}(b)].
We focus on the avoided crossings between energy levels directly connected to the quantum phase slip amplitudes, and measurable via microwave spectroscopy.
We provide analytical expressions, backed by numerics, that capture the entire crossover between $2\pi$- and $4\pi$-dominated regimes near the resonance, as well as the regime away from resonance.

We also show that, when the inductive energy of the loop becomes much smaller than the Josephson plasma frequency, the circuit is well described by a low-energy theory dual to that of a topological superconducting island.
The duality we uncover is distinct from the known duality between a superconducting loop and a superconducting island~\cite{mooij2006}.
Indeed, it includes an additional degree of freedom: the \emph{fluxon parity} of the loop (i.e. the parity of $m$), which we show to be dual to the \emph{fermion parity} of the island.
Similar to fermion parity states encoded non-locally in Majorana zero modes, states with opposite fluxon parity have disjoint support in phase and provide a two-fold quasi-degeneracy to the energy spectrum; thus, they become an attractive degree of freedom to encode qubit states~\cite{kalashnikov2020bifluxon,douccot2012physical}.
We discuss the implications of our findings for the design of protected qubits~\cite{douccot2012physical,gyenis2021moving} in the concluding section.

\section{Model}

Given a capacitance $C$ and an inductance $L$, the inductively shunted junction of Fig.~\ref{fig:layout}(a) is described by the quantum Hamiltonian \cite{koch2009charging}:
\begin{equation}\label{eq:hamiltonian}
\hat{H} = 4 E_c \hat{n}^2 + \tfrac{1}{2} E_L (\hat{\phi}+\phi_\textrm{ext})^2 +V(\hat{\phi})\,,
\end{equation}
where $E_c = e^2/2C$ and $E_L=(\Phi_0/2\pi)^2/L$.
The parameter $\phi_\textrm{ext}= 2\pi\Phi/\Phi_0$ gives the applied flux $\Phi$ through the inductive loop  in units of the flux quantum $\Phi_0=h/2e$.
The Cooper pair number $\hat{n}$ and phase $\hat{\phi}$ are conjugate variables satisfying $[\hat{\phi},\hat{n}]=i$.

\begin{figure}[t!]
    \centering
    \includegraphics[width=\columnwidth]{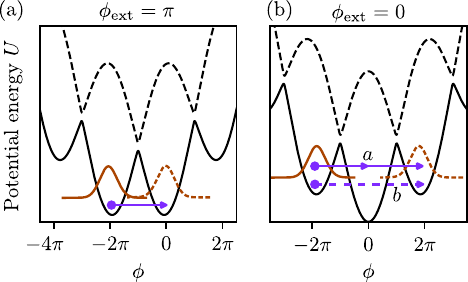}
    \caption{Potential landscape of the model of Eq.~\eqref{eq:hamiltonian}. We depict the two branches of the potential energy $U(\phi) = \tfrac{1}{2}E_L(\phi+\phi_\textrm{ext})^2\pm E_A(\phi)$. 
    (a) When the external flux is equal to half a flux quantum, fluxons are localized around the Josephson potential minima at $\phi=0, 2\pi$ (wave functions shown in orange). Fluxons can tunnel between the minima via a $2\pi$ quantum phase slip (purple arrow).
    (b) When the external flux is zero, fluxons localized around $\phi=\pm 2 \pi$ can tunnel via $4\pi$ quantum phase slips. Because of the second branch of the potential, the $4\pi$ quantum phase slips can follow two interfering paths labeled $a$ and $b$ (solid and dashed arrows), as described in the text.}
    \label{fig:landscape}
\end{figure}

The potential term $V(\hat{\phi})$ gives the Josephson energy, which for a tunnel junction would be the familiar $-E_J\cos\hat\phi$.
For the case in which Josephson coupling is mediated by an isolated energy level, as in Fig.~\ref{fig:layout}(b), a minimal model for the potential is:
\begin{align}\label{eq:potential}
V(\hat\phi) &= - \Gamma\,\cos(\hat\phi/2) \tau_x- \delta\Gamma\,\sin(\hat\phi/2) \tau_y - \epsilon_r \tau_z\,.
\end{align}
Here, the Pauli matrices $\tau_x, \tau_y, \tau_z$ act on the two-level system corresponding to the resonant level being empty or doubly occupied; $\Gamma= \Gamma_1 + \Gamma_2$ and $\delta\Gamma=\Gamma_1-\Gamma_2$ are the sum and difference of the $2e$ tunneling rates $\Gamma_1$ and $\Gamma_2$ between the two leads and the resonant level; and finally $\epsilon_r$ is the energy of the resonant level [see Fig.~\ref{fig:layout}(b)].
This model for the Josephson resonant level has been discussed in Refs.~\cite{meng2009self,kurilovich2021microwave,vakhtel2023quantum}. 
Among other things, these works discuss the role of a charging energy of the resonant level, as well as the effect of additional transport channels and the continuum part of the density of states; all elements which we do not include in our work for simplicity.

The potential in Eq.~\eqref{eq:potential} also applies to the bifluxon circuit deep in the charging regime of the middle island~\cite{kalashnikov2020bifluxon}, but parameters have a slightly different meaning: $\Gamma_1$ and $\Gamma_2$ are Josepshon energies of two tunnel junctions, and $\epsilon_r$ is the energy difference between two even-parity charge states of the superconducting island.

\begin{figure*}[t!]
    \centering
    \includegraphics[width=\textwidth]{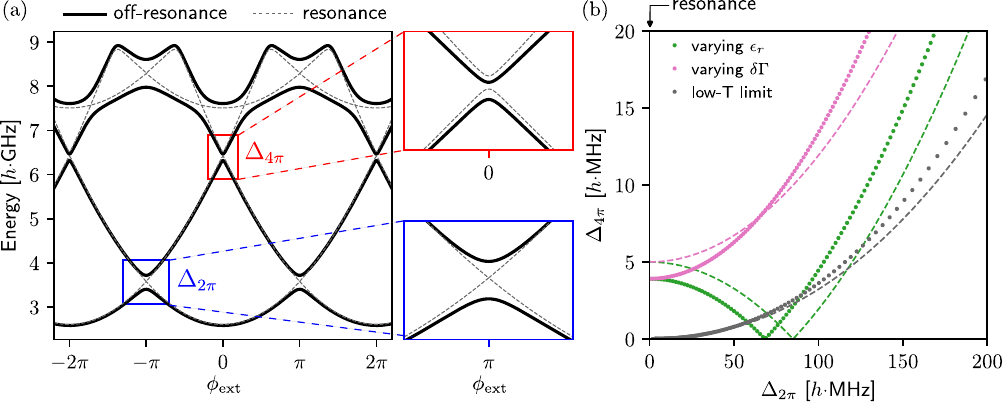}
    \caption{(a) Energy spectrum as a function of flux, $\phi_\textrm{ext}$. The blue and red insets zoom in on the avoided crossings due to $2\pi$ and $4\pi$ quantum phase slips respectively (the vertical span of the insets is $1$~GHz). The energies are computed numerically from Eq.~\eqref{eq:hamiltonian}, with $E_c/h=2.5$ GHz, $E_L/h=0.25$ GHz, $\Gamma/h=5$ GHz, $\delta\Gamma/h=\epsilon_r=0.5$ GHz. These parameters correspond to a reflection coefficient $\abs{r}=0.14$. The dashed gray lines illustrate the resonant case in which $\epsilon_r=\delta\Gamma=0$ and so $r=0$. (b) Comparison of the avoided crossings $\Delta_{2\pi}$ and $\Delta_{4\pi}$ when sweeping system parameters. For all curves, we fix  $E_c/h=2.5$ GHz and $E_L/h=0.25$ GHz. The pink and green data show results obtained approaching resonance in two different ways. In both cases we set $\Gamma/h=10$ GHz. In green, $\epsilon_r/h$ is varied between 0 and 1 GHz, with $\delta\Gamma=0$. In pink, $\delta\Gamma/h$ is varied instead between 0 and 1 GHz, with $\epsilon_r=0$. For both curves, $\abs{r}\approx 0.1$ on the right side of the plot, and tends towards $0$ on the left side of the plot, where $\delta\Gamma=\epsilon_r=0$ and $\Delta_{2\pi}$ vanishes. Dots are computed numerically by diagonalizing Eq.~\eqref{eq:hamiltonian}, while dashed lines are obtained from the WKB result of Eqs.~\eqref{eq:dE_2pi} and \eqref{eq:dE_4pi}. 
    The grey dots show the low-transparency scaling obtained from numerical diagonalization of Eq.~\eqref{eq:hamiltonian} with $V(\hat\phi)=-E_J\cos\hat\phi$, varying $E_J/h$ between $10$ and $40$ GHz. The dashed line corresponds to the $T\ll1$ limit of Eqs.~\eqref{eq:dE_2pi} and \eqref{eq:dE_4pi}, with the correspondence $E_J=\Gamma_AT/4$.}
    \label{fig:splittings}
\end{figure*}

Fluxonium devices are typically operated in a parameter regime such that there is approximately one bound state in each of the local minima of the modulated potential 
of Eq.~\eqref{eq:hamiltonian}~\cite{manucharyan2009fluxonium}.
These bound states are fluxons with a parabolic energy dispersion $\approx \tfrac{1}{2}E_L(2\pi m + \phi_\textrm{ext})^2$~[see Fig.~\ref{fig:splittings}(a)], and become degenerate for particular values of $\phi_\textrm{ext}$.
At the degeneracy points, quantum phase slips create coherent superpositions of fluxons.

In particular, at $\phi_\textrm{ext} = \pi$ the potential landscape is a degenerate double well for fluxons with $m=0$ and $m=-1$, which couple via $2\pi$ quantum phase slips [see Fig.~\ref{fig:landscape}(a)].
At $\phi_\textrm{ext}=0$, instead, fluxons with $m=\pm 1$ occupy degenerate minima symmetrically placed around $\phi=0$, and are coupled by $4\pi$ quantum phase slips [see Fig.~\ref{fig:landscape}(b)].
When $V(\phi)=-E_J\cos\phi$, the $4\pi$ quantum phase slips have a much smaller amplitude than $2\pi$ ones, since they are a higher-order process involving two $2\pi$-slips \cite{koch2009charging}.

This is not necessarily the case for the Josephson resonant level [Eq.~\eqref{eq:potential}], because of the presence of a second branch corresponding to an excited Andreev pair in the junction~\cite{bretheau2013exciting}. 
Indeed, the matrix-valued potential $V(\hat{\phi})$ has eigenvalues $\pm E_A$, with
\begin{equation}\label{eq:EA}
E_A = \Gamma_A \sqrt{\cos^2(\phi/2) + \abs{r}^2 \sin^2(\phi/2)}\,,
\end{equation}
where
\begin{equation}
\Gamma_A = \sqrt{\Gamma^2 + \epsilon_r^2}\,,
\end{equation}
and
\begin{equation}
r = \frac{\epsilon_r + i \delta\Gamma}{\Gamma_A}
\end{equation}
is the reflection amplitude of the junction.

The excited energy branch $+E_A$ is shown as a black dashed line in Fig.~\ref{fig:landscape}(a,b).
The relevant feature of Eq.~\eqref{eq:EA} is an avoided crossing of magnitude $\abs{r}\Gamma_A$ at $\phi=\pm\pi, \pm 3\pi, \dots$.
In the next section we show that in the limit $r\to0$, when the branches cross, the amplitude of $2\pi$ phase slips vanishes. The system thus enters the regime in which $4\pi$ phase slips are dominant.

\section{Wentzel–Kramers–Brillouin (WKB) analysis}
\label{sec:wkb}

An observable consequence of quantum phase slips are avoided crossings in the flux dependence of the energy spectrum of the circuit, see Fig.~\ref{fig:splittings}.
There, $\Delta_{2\pi}$ is the splitting of the crossing between states with $m = 0$ and $m = -1$ at $\phi_\textrm{ext}=\pi$; it originates from $2\pi$ phase slips. $\Delta_{4\pi}$ is the splitting of the crossing between states with $m = -1$ and $m = 1$ at $\phi_\textrm{ext}=0$; it originates from $4\pi$ phase slips.
The magnitude of these avoided crossings can be computed using the WKB method~\cite{landau2013quantum}, with calculations similar to the one described in detail in Ref.~\cite{vakhtel2023quantum}.
One must perform separate calculations to determine $\Delta_{2\pi}$ and $\Delta_{4\pi}$, respectively using the two potential landscapes at $\phi_\textrm{ext}=\pi$ and $\phi_\textrm{ext}=0$ [Fig.~\ref{fig:landscape}(a,b)].
In both cases, the presence of a second branch of the potential crucially modifies the WKB tunneling amplitude under the barrier separating different local minima~\cite{averin1999coulomb,ivanov1998coulomb,averin1999quantum,pikulin2019coulomb,vakhtel2023quantum}.

In this section, we discuss the implications of this fact using a WKB calculation appropriate for the parameter regime typical of fluxonium qubits, in particular with respect to the value of $E_L$.
In the next section, the results are generalized to arbitrarily low values of the inductive energy.

For the $2\pi$ quantum phase slips at $\phi_\textrm{ext}=\pi$, under validity conditions discussed at the end of the Section, one obtains
\begin{equation}\label{eq:dE_2pi}
\Delta_{2\pi} = w(r)\,\omega_p \left(\frac{b_0^2 \omega_p}{2\pi E_c}\right)^{1/2} \exp\left(-b_1\frac{\omega_p}{E_c}+b_2\frac{E_L}{\omega_p}\right)\,.
\end{equation}
where
\begin{equation}
\omega_p = \sqrt{2T\Gamma_A E_c}\,,\quad
T=1-\abs{r}^2\,,
\end{equation}
and $b_0, b_1, b_2$ are numerical coefficients which depend smoothly on the transmission probability $T$. They are given in Appendix~\ref{app:coefficients}. 
The pre-factor $w(r)$ depends on the reflection coefficient $r$ via an adiabaticity parameter $\lambda$:
\begin{equation}\label{eq:prefactor_w}
w(r)=\sqrt{\frac{2 \pi}{\lambda}} \frac{e^{-\lambda} \lambda^\lambda}{\Gamma(\lambda)}\,,\quad\lambda=\frac{\abs{r}^2}{4} \frac{\Gamma_A}{\Gamma}\sqrt{\frac{\Gamma_A}{E_c}}\,.
\end{equation}
with $\Gamma(\lambda)$ the gamma function evaluated at $\lambda$, not to be confused with tunneling rates.
The amplitude $w$ vanishes when $r\to 0$, making the fluxon bound states degenerate at $\phi_\textrm{ext}=\pi$.
The parameter $\lambda$ sets the scale for the crossover into the degenerate regime: the suppression of $\Delta_{2\pi}$ takes place when $\lambda \ll 1$, namely when $\abs{r}^2\ll \sqrt{E_c/\Gamma}$, while $w\approx 1$ in the opposite limit $\lambda \gg 1$.
The mechanism behind the suppression is the imaginary-time Landau-Zener transition across the avoided crossing~\cite{averin1999coulomb}.

The WKB calculation of $\Delta_{4\pi}$ is more delicate, because
there are two tunneling paths between the minima at $\phi = \pm 2\pi$, labeled $a$ and $b$ in Fig.~\ref{fig:landscape}(b).
They differ by the branch of the potential that they take between the two avoided crossings at $\phi=\pm\pi$.
Path $a$ takes place via the lower branch of the potential.
It consists of the sequence of two $2\pi$ phase slips, passing through a classically available region around $\phi=0$.
Path $b$, instead, takes place via the excited branch of the potential and passes through a single $4\pi$-wide tunneling barrier.

Notably, the two contributions interfere.
The interference phase is that of the reflection amplitude $r=\abs{r}e^{i\alpha}$, which distinguishes the path going through the avoided crossings from the one which does not.
The sensitivity of energy levels to the phase acquired at the avoided crossing is akin to the Landau-Zener-St\"{u}ckelberg interference~\cite{shevchenko2010landau}.
The final result for the energy splitting takes the form
\begin{equation}\label{eq:dE_4pi}
\Delta_{4\pi} = \sqrt{\Delta_a^2+\Delta_b^2 - 2\, \Delta_a\,\Delta_b\,\cos(2\alpha)}\,.
\end{equation}
Here, $\Delta_a$ is the contribution due to the sequence of two $2\pi$ phase slips. It takes the form:
\begin{equation}
    \Delta_a = \frac{\Delta_{2\pi}^2}{4\pi^2 E_L}\,\left( \frac{b_0^2\omega_p }{2E_c}\right)^{2\pi^2E_L/\omega_p}
\end{equation}
where $\Delta_{2\pi}$ is the same as given in Eq.~\eqref{eq:dE_2pi}. Note that this contribution vanishes when $r\to 0$.
On the other hand, $\Delta_b$ is the amplitude of a direct $4\pi$ quantum phase slip.
It does not vanish at resonance, and is given by
\begin{align}\label{eq:Delta_b}
\Delta_b = \omega_p\,\left(\frac{b_0^2 \omega_p}{2\pi E_c}\right)^{1/2} \exp\left(-b_3\sqrt{\frac{\Gamma_A}{E_c}} + b_4\frac{E_L}{\omega_p}+ b_5\right) 
\end{align}
with $b_3, b_4, b_5$ another three coefficients smoothly depending on $T$, also given in Appendix~\ref{app:coefficients}.

The results of Eq.~\eqref{eq:dE_2pi} and \eqref{eq:dE_4pi} are illustrated in Fig.~\ref{fig:splittings}.
The parametric plot of $\Delta_{4\pi}$ versus $\Delta_{2\pi}$ shows that, close to resonance, $\Delta_{2\pi}$ vanishes and $\Delta_{4\pi}$ remains finite.
The $4\pi$-dominated regime is approached differently depending on whether the junction is tuned to resonance by varying $\delta\Gamma$ or by varying $\epsilon_r$.
When $\delta\Gamma \neq 0$, $\alpha=\pi/2$ in Eq.~\eqref{eq:dE_4pi}, and so $\Delta_a$ and $\Delta_b$ can never cancel out.
When $\epsilon_r \neq 0$, $\alpha=0$, and so complete cancellation ($\Delta_{4\pi}=0$) occurs at the value of $\epsilon_r$ such that $\Delta_a=\Delta_b$.

\begin{figure*}[t!]
\includegraphics[width=\textwidth]{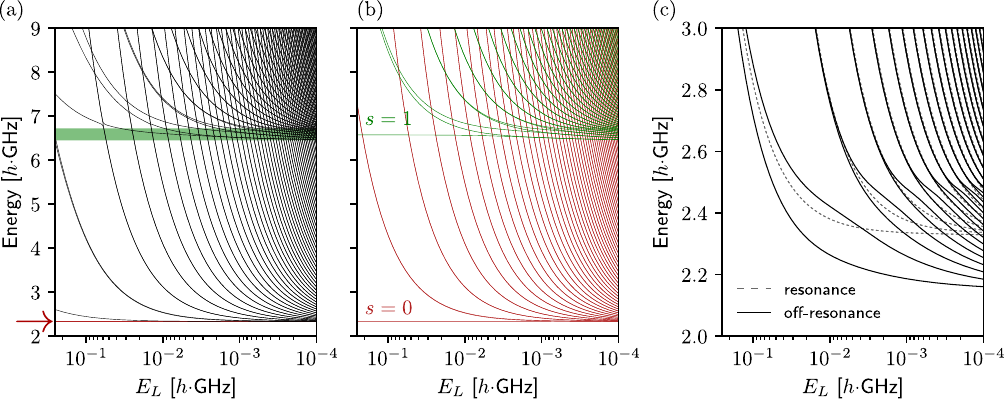}
    \caption{Energy spectrum as a function of decreasing inductive energy $E_L$. (a) Energy levels determined from direct numerical diagonalization of Eq.~\eqref{eq:hamiltonian}; the parameters are $E_c/h=2.5$ GHz, $\Gamma/h=5$ GHz, $\phi_\textrm{ext}=0$, $\epsilon_r/h=10$ MHz and $\delta\Gamma=0$, corresponding to $r\approx 0.002$, very close to resonance. As $E_L\to 0$, the energy levels tend to fill the areas shaded in red and green. These correspond to the energy bands defined in Eq.~\eqref{eq:energybands} for $s=0, 1$. The bandwidth of the $s=0$ band is barely resolvable at about 16 MHz and so it is also indicated by the red arrow. 
    (b) Result of the numerical diagonalization of the effective Hamiltonian $H_s$ of Eq.~\eqref{eq:Hs}, separately for $s=0, 1$.
    The quantum phase slip amplitudes used in the effective Hamiltonian are $A_0\approx 2.8$~MHz, $B_0\approx 6.6$~MHz; and $A_1\approx 7.8$~MHz, $B_1\approx 133$~MHz.
    While the effective spectrum in (b) faithfully reproduces the clustering of energy levels into bands, it does not capture avoided crossings in (a) that originate from the inter-band couplings. (c) Low-lying energy levels computed for the $s=0$ band at $\phi_\textrm{ext}=\pi$ both on resonance ($\delta\Gamma=\epsilon=0$) and off-resonance ($\delta\Gamma=\epsilon_r=0.5$~GHz, i.e., $\abs{r}\approx0.14$). The low-energy effective parameters are $A_0=0$ and $B_0\approx 6.6$~MHz for the resonant case, and $A_0\approx 160$~MHz and $B_0\approx 7.3$~MHz for the off-resonant case. The panel illustrates the different degeneracy of energy levels that is observed in the two cases: degenerate doublets in the resonant case split off-resonance due to $2\pi$ quantum phase slips.}
    \label{fig:low_EL}
\end{figure*}

Eqs.~\eqref{eq:dE_2pi} and \eqref{eq:dE_4pi} are valid when $E_c \ll \Gamma_AT/4$, $E_L \ll \omega_p$, and $\max(\Delta_{2\pi}, \Delta_{4\pi}) \ll E_L$, and apply only to the splitting of fluxons belonging to the lowest harmonic level in the relevant potential minima.
The first condition is required for the validity of the semiclassical WKB approach.
The second condition guarantees that we can disregard fluxons originating from the other harmonic levels inside the wells.
Finally, the third condition allows us to ignore the presence of the higher-energy minima of the potential energy.
The assumed hierarchy of energy scales is in line with experimentally reported parameters of fluxonium devices~\cite{manucharyan2009fluxonium,earnest2018realization,zhang2021universal}, with better accuracy in the ``heavy'' regime $E_c \ll \Gamma_AT/4$~\cite{earnest2018realization,zhang2021universal}.

In Eqs.~\eqref{eq:dE_2pi} and \eqref{eq:Delta_b} we include contributions to the WKB exponent proportional to the small parameter $E_L/\omega_p$.
These contributions originate from the lifting of the energy minima of the periodic potential $V(\phi)$, as well as the change in the WKB momentum due to the $E_L$ term.
Although they are sub-leading contributions to the WKB integrals, and are subtle to compute, we find that they are important for the agreement with numerical calculations in the parameter regime of the aforementioned experiments, such as the parameters used in Fig.~\ref{fig:splittings}.

The analytical results in this Section extend those presented for the same Hamiltonian in Ref.~\cite{kalashnikov2020bifluxon}, which focused on the resonant point $r=0$, since they provide the behavior of $\Delta_{2\pi}$ and $\Delta_{4\pi}$ as the system is tuned across the resonance.
Furthermore, as long as $E_c \ll \Gamma_AT/4$, Eqs.~\eqref{eq:dE_2pi} and \eqref{eq:dE_4pi} remain valid also in the low-transparency regime $T\ll 1$, away from resonance.
In fact, in the limit $T\ll 1$, Eq.~\eqref{eq:dE_2pi}
and \eqref{eq:dE_4pi} match exactly the results of an equivalent WKB calculation done with the tunnel junction potential $V(\hat\phi)=-E_J\cos\hat\phi$, provided that one sets $E_J=\Gamma_A T/4$ so that $\omega_p=\sqrt{8E_JE_c}$.
In this off-resonant regime one always has $\Delta_{4\pi}\ll \Delta_{2\pi}$, as shown by the gray lines in Fig.~\ref{fig:splittings}. 

\section{Duality with a topological superconducting island}
\label{sec:duality}

We now ask what happens to the low energy spectrum when $E_L$ is lowered, so that the assumption $E_L \gg \max(\Delta_{4\pi},\Delta_{2\pi})$ behind the results from the last section is violated and the discussed eigenstates are delocalized over more minima.

The scaling of the energy spectrum of Eq.~\eqref{eq:hamiltonian} towards the limit $E_L\to 0$ is shown in Fig.~\ref{fig:low_EL}.
In the limit $E_L\ll \omega_p$, as more and more local minima of the potential appear at energies below $\omega_p$, we observe the condensation of bands of narrowly spaced energy levels.
We now derive an effective Hamiltonian appropriate to describe this regime, via similar steps as those described in Ref.~\cite{koch2009charging} for the standard fluxonium Hamiltonian.
The derivation will establish the duality with the topological superconducting island mentioned in the introduction.

To begin with, when $E_L\ll \omega_p$, it becomes convenient to write the Hamiltonian~\eqref{eq:hamiltonian} in the eigenbasis of its $E_L \rightarrow 0$ limit. The eigenfunctions can be represented in the following way:
\begin{equation}
    \Psi_{ns}(\phi)=e^{-i n \phi} u_{ns}(\phi) \equiv\langle\phi \mid n, s\rangle .
\end{equation}
Here, $s$ is an integer number that refers to a band index and $n$ is a continuous variable, $n \in [0,1)$.
By substitution into~\eqref{eq:hamiltonian}, the spinor wave functions $u_{ns}(\phi)$ satisfy a transmon-like equation:
\begin{align}
&[4 E_c (-i \partial_\phi-n)^2+V(\phi)]\,u_{ns} = E_s(n)\,u_{ns},
\end{align}
with the boundary condition that was derived in Ref.~\cite{vakhtel2023quantum}:
\begin{equation}
u_{ns}(\phi+2\pi) = \tau_z u_{ns}(\phi)\ .
\end{equation}
Note that $u_{ns}$ are defined on the circle $\phi \in [0,2\pi)$ and, at a fixed $n$, they form an orthonormal basis with respect to the band index $s$. This ensures that $\Psi_{ns}(\phi)$, which are functions of a non-compact phase, form an orthonormal basis with different $s$ and $n$.

This eigenvalue problem was analyzed in Ref.~\cite{vakhtel2023quantum}, where we showed that the eigenspectrum takes the form:
\begin{equation}\label{eq:energybands}
E_s(n) = \epsilon_s + A_s\cos(2\pi n + \alpha_s) + B_s\cos(4\pi n + \beta_s)\,.
\end{equation}
Here, $A_s$ and $B_s$ are the $2\pi$ and $4\pi$ quantum phase slip tunneling amplitudes for the periodic potential $V(\phi)$, and $\alpha_s$ and $\beta_s$ are associated phase shifts.
The bands are harmonically spaced, $\epsilon_s\approx \omega_p(s+\tfrac{1}{2})$, while $A_s$ and $B_s$ are exponentially small in $\omega_p/E_c$.
Detailed expressions as a function of $E_c$, $\Gamma$, $\delta\Gamma$ and $\epsilon_r$ are derived in Ref.~\cite{vakhtel2023quantum} and restated in Appendix~\ref{app:effective_parameters}. 
The simple form above for the energy bands was derived via the WKB method.
It is accurate for $E_c \ll \Gamma_A T/4$ and for low-lying bands.

For the lowest band, the parameters $A_0$ and $B_0$ are closely connected to the quantities $\Delta_{2\pi}$ and $\Delta_{4\pi}$ computed in the previous section.
In particular, $A_0$ can be identified with the limit $E_L/\omega_p\to 0$ of $\Delta_{2\pi}$ in Eq.~\eqref{eq:dE_2pi}, but the same is not true for $B_0$, since in Eq.~\eqref{eq:dE_4pi} the ratio $\Delta_{2\pi}/E_L$ appears as well (i.e., both $A_0$ and $B_0$ contribute to $\Delta_{4\pi}$).
We have verified numerically that the low-energy spectrum of the $s=0$ band, discussed in more detail below, matches the expressions for the energy splittings given in Eqs.~\eqref{eq:dE_2pi} and \eqref{eq:dE_4pi}.
This is true provided $E_L$ is low enough to neglect the sub-leading $E_L/\omega_p$ terms in those equations, but large enough so that $E_L\gg\max(\Delta_{2\pi},\Delta_{4\pi})$ as required in the previous section.

In the basis $\ket{n,s}$, the phase operator is represented as $\hat{\phi}= - i\partial_{n} - \hat{\Omega}$.
It couples different bands only via the connection matrix elements $\Omega_{s s'}$:
\begin{align}
\begin{aligned}
\bra{n,s}\hat{\Omega}\ket{n^{\prime},s^{\prime}} =   \delta (n-n^{\prime}) \Omega_{s s'}(n) \\ \Omega_{s s'}(n) = i \int^{2\pi}_{0} u_{n s}^{\dagger}\partial_{n} u_{n s^{\prime}} d\phi
\end{aligned}
\end{align}
These can be evaluated in the same limit where~\eqref{eq:energybands} was calculated:
\begin{equation}
   \Omega_{s s^{\prime}}(n) \approx - \left(\frac{8 E_c}{\Gamma_A T}\right)^{1 / 4}\left(\sqrt{s} \delta_{s^{\prime}, s+1}+\sqrt{s+1} \delta_{s^{\prime}, s-1}\right) .
\end{equation}
The interband couplings can be neglected for $E_c \ll \Gamma_A T/4$.
Therefore, the original Hamiltonian of Eq.~\eqref{eq:hamiltonian} separates into blocks labelled by the band index $s$:
\begin{equation}\label{eq:Hs}
H_s = \tfrac{1}{2}E_L(-i\partial_n+\phi_\textrm{ext})^2 + E_s(n)\,.
\end{equation}
It must be solved with the periodic boundary conditions $\psi_s(n+1)=\psi_s(n)$.
The eigenvalues of this block-diagonal Hamiltonian, shown in the right panel of Fig.~\ref{fig:low_EL}, compare favorably to the numerical solution of the full Hamiltonian, Eq.~\eqref{eq:hamiltonian}, shown in the left panel of  Fig.~\ref{fig:low_EL}.

The fluxon states localized around minima $\phi=2\pi m$ with integer $m$ are related to $\ket{n,s}$ via the Fourier transform:
\begin{align}\label{eq:fluxons}
    \ket{2\pi m, s} = \int_0^1 d n e^{2 \pi i m n}\ket{n,s}.
\end{align}
It is easy to see that, at resonance, $A_s$ vanishes and the parity of $m$ becomes conserved.
With this in mind, we introduce in lieu of $\ket{n,s}$ a new basis $\ket{n, \sigma, s}$ endowed with a spin-like degree of freedom related to the fluxon parity:
\begin{align}
    \ket{n, \up, s} &= \frac{\ket{n,s}+\ket{n +1/2,s}}{\sqrt{2}}\,, \\
    \ket{n, \down, s} &=\frac{ \ket{n,s}-\ket{n +1/2,s}}{\sqrt{2}}\,,
\end{align}
with $n\,\in\,[0,1/2)$. In terms of these basis states, \begin{equation}
\ket{2\pi m,s} = \sqrt{2}\int^{1/2}_{0} d n \ket{n,\sigma,s} e^{2\pi i m n}\,,
\end{equation}
where $m$ is even for $\sigma=\,\uparrow$ and odd for $\sigma=\,\downarrow$.

\begin{figure}[t!]
    \includegraphics[width=\columnwidth]{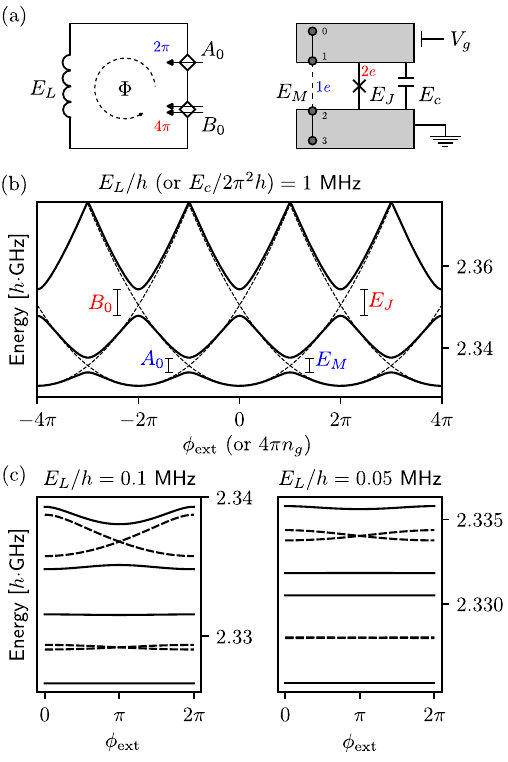}
    \caption{
    (a) Schematic illustration of the duality between a supercondcting loop (left) with $2\pi$ and $4\pi$ phase slip elements $A_0$ and $B_0$ and a topological superconducting island (right) with $1e$ and $2e$ tunnel couplings $E_M$ and $E_J$. The four grey dots on the right represent four Majorana zero modes, two on the island and two on the ground. $\Phi$ and $V_g$ are the flux and voltage applied to the loop and island, corresponding to the tuning parameters $\phi_\textrm{ext} = 2\pi\,\Phi/\Phi_0$ and $n_g = CV_g/2e$.
    (b) Dispersion of the three lowest energy levels of the circuit as a function of flux, obtained from diagonalization of the $s=0$ band Hamiltonian of Eq.~\eqref{eq:Hs}. We set $E_c/h=2.5$ GHz and $\Gamma/h=5$ GHz, and $\epsilon_r/h=\delta\Gamma/h=10$ MHz. In these conditions, the quantum phase slip parameters of Eq.~\eqref{eq:Hs} are $A_0/h\approx 3.3$ MHz and $B_0/h\approx 6.7$ MHz. The dashed parabolas are the energies of uncoupled fluxons, which are dual to uncoupled charge states. Labels relate avoided crossings to model parameters on either side of the duality.
    (c) Flux dispersion of the energy levels of the circuit for lower values of $E_L$, illustrating the ``transmon'' regime. The solid lines are obtained for the same parameters as in panel (b), while the dashed lines are obtained at resonance: $\epsilon_r=\delta\Gamma=0$. In this case, $A_0=0$ and energy levels gather in almost degenerate doublets. In panel (b), $B_0 / (2\pi^2 E_L) \approx 0.3$, while in panel (c) $B_0/(2\pi^2 E_L) \approx 3.3$ and $6.6$. Note that the vertical energy scale changes between plots, following the reduction in $E_L$.}
    \label{fig:duality}
\end{figure}

The Hamiltonian $H_s$ in this doubled space reads
\begin{align}\nonumber
H_s &= \tfrac{1}{2}E_L(i\partial_n-\phi_\textrm{ext})^2 + A_{s}\sigma_x\cos(2\pi n + \alpha_s) \\\label{eq:Hs_with_spin}
&+ B_{s}\cos(4\pi n + \beta_s)\,+ \epsilon_s\,.
\end{align}
The Pauli matrices act on the spin-like degree of freedom and the boundary conditions in the halved Brillouin zone become twisted:
\begin{equation}\label{eq:twisted_bc}
\psi_s(n+\tfrac{1}{2}) = \sigma_z \psi_n(n)\,.
\end{equation}
Although Eq.~\eqref{eq:Hs_with_spin} is just a re-writing of Eq.~\eqref{eq:Hs}, it illuminates the fact that the low-energy description is precisely dual to that of a superconducting island shunted to ground by a topological Josephson junction with coupled Majorana zero modes~[see Fig.~\ref{fig:duality}(a)].
The Hamiltonian of such an island is~\cite{heck2011,ginossar2014microwave,ulrich2014,yavilberg2015,rodriguezmota2019,svetogorov2020,karki2024,pino2024}
\begin{align}
\nonumber
H_M &= 4E_c(i\partial_\phi - n_g)^2 + E_M\,i\gamma_1\gamma_2\,\cos(\phi/2)\\
&- E_J \cos\phi\,.
\label{eq:Hisland}
\end{align}
Here, the first term is the charging energy of the island, $n_g$ is the induced charge in units of $2e$, $E_J$ represents standard Cooper pair tunneling, and the last term represents single-charge tunneling due to the Majorana zero modes $\gamma_1$ and $\gamma_2$ coupled across the topological junction (the fractional Josephson effect).
Note that there are four Majorana zero modes in the model, with $\gamma_0$ and $\gamma_1$ located on the island and $\gamma_2$ and $\gamma_3$ located on the ground plane (see Fig.~\ref{fig:duality}).
Although only $\gamma_1$ and $\gamma_2$ appear in the Hamiltonian, the boundary condition for Eq.~\eqref{eq:Hisland} depends on the fermion parity operator of the island $i\gamma_0\gamma_1$:
\begin{equation}\label{eq:twist}
\psi(\phi+2\pi) = (-1)^p\,\psi(\phi)\,,
\end{equation}
with $p=(1-i\gamma_0\gamma_1)/2 = 0$ or $1$ if the parity is even or odd. The operator $i\gamma_0\gamma_1$ appearing in the boundary condition
anticommutes with $i\gamma_1\gamma_2$ appearing in the Hamiltonian, just like the fluxon parity $\sigma_z$ entering the boundary condition of Eq.~\eqref{eq:twisted_bc} anticommutes with $\sigma_x$.

As illustrated in Fig.~\ref{fig:duality}(a) and (b), the duality is established via the following correspondences: $\phi \leftrightarrow 4\pi n$, $\phi_\textrm{ext} \leftrightarrow 4\pi n_g$, $E_c \leftrightarrow 2\pi^2 E_L$, $E_J \leftrightarrow B_s$, $E_M \leftrightarrow A_s$.
The operator $i\gamma_1\gamma_2$ changes the fermion parity of the island, just like the operator $\sigma_x$ changes the fluxon parity.
The phase shifts $\alpha_s$ and $\beta_s$ can be included in the correspondence by adding a relative phase between the $E_M$ and $E_J$ terms, which could arise for instance in a superconducting quantum interference device (SQUID) configuration.

It follows from the duality that, in the limit of low $E_L$, the flux dispersion of the energy levels of the circuit is equivalent to the charge dispersion of the energy levels of a superconducting island governed by Eq.~\eqref{eq:Hisland}.
With due care, results available in the literature for the latter system become therefore applicable to the inductive loop as well.
This includes, for instance, the existence of a supersymmetric spectrum at a specific value of the system parameters~\cite{ulrich2014}.

We illustrate the salient aspects of the duality in Fig.~\ref{fig:low_EL}(c), focusing on the lowest energy levels of the $s=0$ band when $\phi_\textrm{ext}=\pi$.
In this case, at resonance, fluxon parity provides a two-fold degeneracy to the energy spectrum of the circuit, which is broken by $2\pi$ quantum phase slips away from resonance.
The flux dispersion of energy levels away from this point is instead shown in Fig.~\ref{fig:duality}: when $2\pi^2E_L \gg B_0$, the circuit is in a ``Cooper-pair box regime'': the energy levels are essentially given by parabolas with small avoided crossings at degeneracy points~[see Fig.~\ref{fig:duality}(b)].
On the other hand, when $2\pi^2 E_L \ll B_0$, the circuit is in a ``transmon regime'' [see Fig.~\ref{fig:duality}(c)], characterized by a flattening of the dispersion of energy levels as a function of flux.
The spacing between these flat energy levels depends on the value of $A_0$. If $A_0=0$, the energy levels become fluxon-parity degenerate doublets at all values of the flux in the limit $E_L\to 0$, with a spacing between doublets $\sim \sqrt{E_LB_0}$ [dashed lines in Fig.~\ref{fig:duality}(c)].
A finite but small $2\pi$ quantum phase slip amplitude splits the doublets by an amount $\approx A_0$ [solid lines in Fig.~\ref{fig:duality}(c)].

\section{Discussion}

\subsection{Observability of the \texorpdfstring{$4\pi$}{4pi}-dominated regime}

The difficulty of measuring directly the $4\pi$-dominated regime occurring at resonance lies in the smallness of $4\pi$ quantum phase slips.
This was the reason, for instance, that the effect of $4\pi$ quantum phase slips was not detected in the transmon experiments of Ref.~\cite{bargerbos2020observation,kringhoj2020suppressed}.
The results of Fig.~\ref{fig:splittings} show that measuring the $4\pi$-dominated regime should be feasible in circuit with typical fluxonium parameters: $E_c/h=2.5$ GHz and $E_L/h=0.25$ GHz.
At perfect resonance, when $\Delta_{2\pi}$ vanishes, $\Delta_{4\pi}/h\approx 5$~MHz if $\Gamma/h\approx 5$~GHz: albeit small, splittings of this magnitude have been detected and exploited in heavy fluxonium circuits~\cite{earnest2018realization,najerasantos2024}.
Larger values of $\Delta_{4\pi}$ can be obtained by decreasing $\Gamma_A/E_c$ (somewhat exiting the domain of validity of our WKB results).

The $4\pi$-dominated regime is narrow: with the parameters of Fig.~\ref{fig:splittings}, one needs $\epsilon_r/\Gamma \lesssim 10^{-3}$ and $\delta\Gamma/\Gamma \lesssim 10^{-3}$ to achieve $\Delta_{2\pi} \lesssim \Delta_{4\pi}$.
For bifluxon circuits, it may be difficult to limit the asymmetry $\delta\Gamma$, which is set by the fabrication of the tunnel junctions~\cite{hertzberg2021laser,takahashi2022uniformity} and cannot be tuned afterwards, unless SQUIDs are added to the design for the purpose.
For semiconductor junctions, instead, a difficulty would be to maintain $\epsilon_r$ and $\delta\Gamma$ in such narrow ranges in the presence of charge noise.
However, we argue that semiconductor junctions present a qualitative advantage relative to the bifluxon: stronger coupling between the weak link region and the superconducting leads can be achieved without sacrificing anharmonicity, namely without compromising the two-level approximation used in the model for the weak link~\cite{bozkurt_josephson_2023}.
As we explain below, the possibility to increase $\Gamma$ without exiting the regime of validity of the model may be beneficial to find a parameter regime which offers more benevolent conditions to observe the $4\pi$-dominated regime.

\subsection{Relation of the duality with previous work on protected qubits}

The duality of Sec.~\ref{sec:duality} is not the first to establish a connection between charge-based and flux-based superconducting circuits.
Notably, Mooij and Nazarov established a duality between the Cooper-pair box and the phase-slip junction~\cite{mooij2006}.
The crucial difference is that, in our case, $2\pi$ quantum phase slips are dual to charge $1e$ tunneling, rather than $2e$ tunneling.
This different mapping means that the duality of Mooij and Nazarov cannot be reobtained simply by disregarding $4\pi$ phase slips.
Namely, even if setting $E_M$ and $A_s$ to zero formally recovers the dual Hamiltonians of Ref.~\cite{mooij2006}, these Hamiltonians in our case act on a different Hilbert space, enlarged by the presence of a degenerate parity degree of freedom.
A conceptually similar duality involving charge and flux degrees of freedom was discussed in Ref.~\cite{ulrich2016}, but it applied to the case of a topological superconducting loop.

The duality of Sec.~\ref{sec:duality} highlights an equivalence between different models of protected superconducting qubits.
Namely, models on both sides of the duality can be cast as a one-dimensional tight-binding model in which the nearest-neighbor hopping ($E_M$ or $A_0$) can become smaller than the next-nearest neighbor hopping ($E_J$ or $B_0$); the hopping represents tunneling of charge or flux depending on the side of the duality.
When the nearest-neighbor hopping is set to zero but the next-nearest neighbor hopping is not, the one-dimensional lattice disconnects in two separate pieces, corresponding to ``even'' and ``odd'' sites of the lattice.
Protected qubits can then be encoded in the parity degree of freedom: parity states are degenerate and have disjoint support.
The degeneracy is broken by the inductive or charging energy, which assigns different energies to even and odd sites, but does not couple them \footnote{When dealing with the Hamiltonian~\eqref{eq:Hisland}, one has to keep in mind that the boundary conditions~\eqref{eq:twist} are twisted. After one makes a gauge transform to change the boundary conditions to periodic, the equilibrium charge $n_g$ in the kinetic term is shifted differently for different parities.}.
With this general picture in mind, it becomes intuitive to see that the duality can be extended to other circuits -- for instance, a transmon with both a $\cos(\phi)$ and a $\cos(2\phi)$ Josephson element~\cite{smith2020}.

\subsection{Implications for qubit protection}

The duality derived in Sec.~\ref{sec:duality} is suggestive for the design of protected qubits.
In the topological superconducting island, the regime $E_M=0$ defines a parity-protected qubit~\cite{hassler2011}: as long as $E_J \gg E_c$, noise acting on the island can neither dephase nor flip the qubit encoded in the fermion parity of the Majorana pair. In our inductive loop, a similar regime corresponds to the resonant condition $A_0=0$ together with the condition $2\pi^2 E_L \ll B_0$~\cite{kalashnikov2020bifluxon}. In this regime, noise in the loop cannot dephase or flip the qubit encoded in the fluxon parity of the loop. The former process is suppressed exponentially in the ratio $\sqrt{8B_0/(2\pi^2E_L)}$.

With the inductive loop also insensitive to charge noise, it appears that, on the theoretical level, the remaining fragility of the protected regime is the fine-tuning needed to establish the resonant condition $\delta\Gamma=\epsilon_r=0$.
Slightly away from resonance, $2\pi$ quantum phase slips couple fluxons of different parity and break the parity protection.
This fine-tuning is problematic in the presence of gate-induced noise influencing the parameters $\delta\Gamma$ and $\epsilon_r$.
This fragility was already noted in the bifluxon proposal of Ref.~\cite{kalashnikov2020bifluxon}, which discussed possible circuit extensions mitigating the problem.
Similar workarounds could be applied to semiconductor junctions, which are also sensitive to this type of noise.

Is there a parameter regime of the model that circumvents this fragility?
The duality suggests a negative answer, as follows.
In the dual model for a topological superconducting island, the topological protection of a fermion parity qubit is spoiled by a non-zero $E_M$.
However, $E_M$ can be pushed towards zero with exponential accuracy if the junction providing it is in the tunneling regime, or pinched-off (or even absent).
At the same time, $E_J$ can be kept large by a different junction in parallel.
Thus, the protected regime is available without the need for fine-tuning circuit parameters.

In the inductive loop, the undesired coupling is the $2\pi$ quantum phase slips amplitude $A_0$ (see Fig.~\ref{fig:duality}). 
Away from the resonant condition, $A_0$ cannot be reduced exponentially while keeping the $4\pi$ phase slips amplitude $B_0$ finite and large, as required by the protected regime.
As both parameters are controlled by the WKB integrals under the potential barrier, reducing $A_0$ (e.g. by increasing the ratio $\omega_p/E_c$) will reduce $B_0$ as well, moving the device away from the protected regime $B_0\gg 2\pi^2 E_L$.
Unlike in the topological island case, the problem cannot be solved by asking for a second junction to provide a large $B_0$ and no $A_0$, because the only way to do so would require fine-tuning this second junction as well.

Finally, the duality illuminates another aspect of the protected regime.
In the topological superconducting island, quasiparticle poisoning would spoil the fermion parity qubit by introducing incoherent bit-flip errors between qubit states~\cite{goldstein2011,budich2012,rainis2012,karzig2021}.
The analogue poisoning processes on the fluxon-based side of the duality are incoherent $2\pi$ phase slips occurring in the inductive loop, e.g.~of thermal origin.
As this source of poisoning may be easier to keep under control, on this point the fluxon-based design seems to have an advantage with respect to its dual -- although quasiparticle poisining \emph{of the resonant level} must still be minimized, as discussed below.

\subsection{Experimental perspectives}

From a practical point of view, an immediate problem with the protected regime of our model is the requirement for extreme smallness of $E_L$: to the best of our knowledge, the current record in the literature stands at $E_L/h\approx 65$~MHz~\cite{pechenezhskiy2020superconducting}, likely higher than what would be needed for the condition $2\pi^2 E_L \ll B_0$.
A related issue is that the level spacing would be in the MHz range, requiring some active cooling to reach the quantum regime at accessible temperatures (milliKelvin scale).
At low values of $E_L$ -- often reached via high-kinetic inductance thin-films with very low Cooper-pair densities -- the occurrence of phase slips across the inductor, neglected here, may also have to be taken into account.

A common strategy to minimize all the problems mentioned so far is to increase the quantum phase slip rates as well as the plasma frequency, essentially trying to maximize both $\Gamma_A$ and $E_c$ while keeping the ratio $\Gamma_A /E_c$ constant and of order one.
Using superconductors with a larger energy gap than Al in the resonant level junction would allow more room to increase $\Gamma_A$ without exiting the tunneling limit. 
It is also essential to minimize the quasiparticle poisoning rate of the quantum dot (which is an analogue of the poisoning events of the Cooper pair box island in the bifluxon \cite{kalashnikov2020bifluxon}), as our model \eqref{eq:potential} assumes even occupation numbers of the Andreev bound state. 

Despite these obstacles, the existence of a protected regime, corroborated by the duality derived in this work, will make it interesting and rewarding to reach the hard-to-reach parameter regime in which the inductive energy becomes much smaller than the quantum phase slip rates.

\acknowledgments{We acknowledge helpful discussions with Vlad Kurilovich, Leonid Glazman, Marcelo Goffman, Christian Urbina, Emmanuel Flurin, and Joan Cáceres. T.V. has received funding from the European Research Council (ERC) under the European Union’s Horizon 2020 research and innovation programme. M.P.V. and A.B. acknowledge financial support from the Dutch Research Council (NWO), with project number 14SCMQ02, and from the Microsoft Quantum initiative. V.F. acknowledges: Research was sponsored by the Army Research Office and was accomplished under Grant Number W911NF-22-1-0053. The views and conclusions contained in this document are those of the authors and should not be interpreted as representing the official policies, either expressed or implied, of the Army Research Office or the U.S. Government. The U.S. Government is authorized to reproduce and distribute reprints for Government purposes notwithstanding any copyright notation herein.}

\appendix 
\section{Definitions of the coefficients}
\label{app:coefficients}

\begin{figure}[t!]
    \includegraphics[width=\columnwidth]{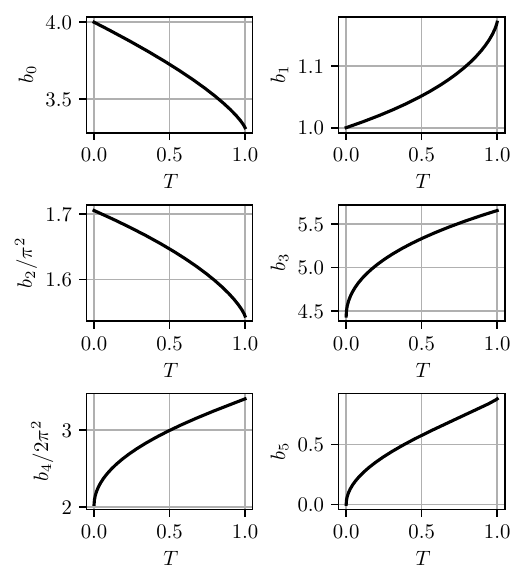}
    \caption{Coefficients $b_0, b_1, b_2, b_3, b_4, b_5$ versus transparency $T$. Their limiting values at $T=0$ are $b_0\rightarrow 4$, $b_1\rightarrow 1$, $b_2\rightarrow 14\zeta(3)$, $b_3\rightarrow\sqrt{2}\pi$, $b_4 \rightarrow 8 \pi G+14\zeta(3)$ where $G = 0.915\dots $ is Catalan's constant, and $b_5  \rightarrow 0$. Their limiting value for $T\to 1$ are $b_0 \rightarrow 8 (\sqrt2-1)$, $b_1 \rightarrow \sqrt{8} (\sqrt2-1)$, $b_2 \rightarrow 15.245\dots$, $b_3 \rightarrow 4\sqrt{2}$, $b_4 \rightarrow  56 \zeta(3)$, $b_5 \rightarrow \arctanh (1/\sqrt{2})$.}
    \label{fig:coefficients}
\end{figure}

In this Appendix, we give the explicit expressions for the coefficients $b_0, b_1, b_2, b_3, b_4, b_5$ used in the paper. We introduce auxiliary definitions first:
\begin{align}
u(\phi) &=E_A(\phi)/\Gamma_A = \sqrt{1-T\sin^2(\phi/2)}\,,\\
\mu(\phi)&=\arcsin\sqrt{\frac{u(\varphi)-\abs{r}}{u(\varphi)+\abs{r}}}\,,\\
h & = \frac{\abs{r}}{(1+\abs{r})\,\sqrt{1-\abs{r}}}\,,\\
k &= \sqrt{\frac{1-\abs{r}}{1+\abs{r}}}\,.
\end{align}
Then $b_0$ and $b_1$ are defined in terms of elliptic integrals of the first and second kind, as follows:
\begin{align}
b_0 &=\lim_{\psi \rightarrow 0}\psi \ e^{\sqrt{2}h \left[2 \Pi(\mu(\psi),k^{-2},k)- (1-\abs{r})F(\mu(0),k) \right]}\,\\
b_1&=\sqrt{8} h \left[-F(\mu(0),k) +2\Pi(\mu(0),1,k)\right]\,.
\end{align}
For the rest of the coefficients, we have:
\begin{align}
b_2 &= \sqrt{\frac{T}{8}}\int^\pi_0 \frac{(\pi^2-\phi^2)d\phi}{\sqrt{1-\sqrt{1-T \cos^2\phi/2}}}.\\
b_3 &= \sqrt{2T}\,b_1 + \int^{\pi}_{0}\sqrt{1+u(\phi)}d\phi\,\\
b_4 &= \sqrt{\frac{T}{8}}\int^{\pi}_{0}\left[\frac{\phi(4\pi-\phi)}{\sqrt{1-u(\phi)}}+ \frac{4\pi^2-\phi^2}{\sqrt{1+u(\phi)}}\right]d\phi\,,\\
b_5 &=\sqrt{\frac{T}{8}}\int^\pi_0\frac{d\phi}{\sqrt{1+u(\phi)}}\,.
\end{align}
The coefficients are plotted against transparency $T$ in Fig.~\ref{fig:coefficients}.

\section{Expressions for the low-energy Hamiltonian parameters}

\label{app:effective_parameters}

The expressions below are given in Ref.~\cite{vakhtel2023quantum}, where $A_s, B_s$ are denoted $\delta^{2e}_s, \delta^{1e}_s$ and $\alpha_s, \beta_s$ are denoted $\beta^{2e}_s,\beta^{1e}_s$ respectively. We state them here for convenience. They have been derived using parabolic cylinder functions near the minima of the Josephson potential. The intermediate expressions \eqref{B1},~\eqref{B2},~\eqref{B3} are different from \cite{vakhtel2023quantum}, but the results for $A_s, \alpha_s$ and $B_s, \beta_s$ are the same after the substitution of ~\eqref{B2} into \eqref{B1},~\eqref{B3}. The $2\pi$-tunneling amplitude and phase for a band $s$ are given by:
\begin{align}\label{B1}
A_s &= \frac{w \omega_p }{z\pi} \,e^{-\tau_s}\,, \ \ \ \alpha_s =\pi(s+1)-\alpha\   \\
z &=\frac{s! e^{s+1/2}}{(s+1/2)^{s+1/2} \sqrt{2\pi}} , 
\end{align}
with $w$ and $\alpha$ as defined in the main text. Here, $\tau_s$ is some WKB integral that can be evaluated to:
\begin{equation}\label{B2}
e^{-\tau_s}=\frac{z\sqrt{2\pi}}{s!}\left(\frac{b_0^2 \omega_p}{4 E_c}\right)^{s+\frac{1}{2}} \exp\left(-b_1 \frac{\omega_p}{E_c}\right)\, . 
\end{equation}
Note that the expression for $A_{s=0}$ coincides with Eq.~\eqref{eq:dE_2pi} when $E_L/\omega_p\to 0$.

For $4\pi$ phase slips, there are two terms contributing to the overall amplitude $B_s$ and phase $\beta_s$, which are defined by the equality:
\begin{align}\label{B3}\nonumber
B_s \cos (4 \pi n+\beta_s)=  \frac{(-1)^{s+1}\omega_p}{\pi z}e^{-\rho_s} e^{-\tau_s} \cos \left(4 \pi n \right) &\\
 +\frac{w^2\omega_p}{2 \pi^2 z^2 } \log \left[\frac{b_0^2 \omega_p}{4 E_c\left(s+\tfrac{1}{2}\right)}\right] e^{-2 \tau_s} \cos \left(4 \pi n-2 \alpha \right).
\end{align}
Here $\rho_s$ is another WKB integral, this time evaluating to
\begin{align}
\rho_s =(b_3-\sqrt{2T}b_1) \sqrt{\frac{\Gamma_A}{E_c}}-b_5\,(2s+1) .
\end{align}

\bibliography{references}

\end{document}